# Seventeen New Variable Stars Detected in Vulpecula and Perseus

**Riccardo Furgoni**

*Keyhole Observatory MPC K48, Via Fossamana 86, S. Giorgio di Mantova (MN), Italy, and AAMN Gorgo Astronomical Observatory MPC 434, S. Benedetto Po (MN), Italy; riccardo.furgoni@gmail.com*

**Abstract**    I report the discovery of seventeen new variable stars in the Northern Sky: three eclipsing (GSC 02129-00759; GSC 02129-00947; GSC 02869-02559), one eruptive (GSC 02856-02521), ten pulsating (2MASS J19305329+2558520; GSC 02129-00537; 2MASS J19323543+2524000; 2MASS J19263580+2616428; HD 275169; GSC 02869-00313; GSC 02869-01981; GSC 02856-01391; GSC 02860-01552; GSC 02856-01465) and three rotating (one of which is suspected) (GSC 02142-01107; GSC 02856-00169; GSC 02865-01593).

## 1. Introduction

Starting in the spring of 2011 the Keyhole Observatory MPC K48, located in S. Giorgio di Mantova (Italy), has been involved in an intense surveying activity mainly focused on new variable star discovery and characterization. Even this year a new observing campaign was conducted in three separate fields located in the constellations of Vulpecula and Perseus for a total of 28 nights, obtaining 2,583 images in the V passband with an overall exposure time of almost 86 hours.

Subsequently, the light curves of all stars in the fields in a magnitude range between 9 V and 14 V have been visually inspected in order to determine the candidate variables, starting with those that have an RMS scatter as function of magnitude higher than normal. When possible the observations were combined with SuperWASP (Butters *et al.* 2010), ASAS-3 (Pojmański 2002), and NSVS (Woźniak *et al.* 2004) datasets to enhance the precision in the determination of the period, magnitude range, and the type of variability.

Considering all the new variables discovered, two are in my opinion the most interesting: GSC 02856-02521, an RS CVn variable with Algol-type eclipses, and GSC 02860-01552, a high-amplitude δ Scuti (HADS) double mode variable.

## 2. Instrumentation used

The data were obtained with a TS Optics APO906 Carbon, a triplet FPL-53 apochromatic refractor with 90-mm aperture and f/6.6 focal ratio; the telescope was also equipped with a field flattener for pin-point stellar images corner-to-corner.

The telescope was positioned at coordinates 45° 12' 33" N, 10° 50' 20" E (WGS84) at the Keyhole Observatory, a roll-off roof structure managed by the author. The pointing was maintained with a Syntha NEQ6 mount guided via a Baader Vario Finder telescope equipped with a Barlow lens capable of bringing the focal length of the system to 636 mm and focal ratio of f/10.5. The guide camera was a Magzero MZ-5 with Micron MT9M001 monochrome sensor. The CCD camera was a SBIG ST8300m with monochrome sensor Kodak KAF8300. The photometry in the Johnson V passband was performed with an Astrodon Photometrics Johnson-V 50-mm round unmounted filter, equipping a Starlight Xpress USB filterwheel.

The camera is equipped with a 1,000× antiblooming: after exhaustive testing it has been verified that the zone of linear response is between 1,000 and 20,000 ADU, although up to 60,000 ADU the loss of linearity is less than 5%. The CCD is equipped with a single-stage Peltier cell $\Delta T = 35 \pm 0.1°$ C which allows the cooling at a stationary temperature.

## 3. Data collection

The observed fields are centered, respectively, at coordinates (J2000) R.A. $19^h 29^m 00^s$, Dec. +25° 50' 00" (Vulpecula), and R.A. $03^h 19^m 00^s$, Dec. +41° 29' 00" and R.A. $03^h 19^m 00^s$, Dec. +42° 46' 00" (both in Perseus). For all, the dimensions are 78' × 59' with a position angle of 360°.

These fields were chosen to maximize the possibility of discovering new variable stars. Their determination was made trying to meet the following criteria:

• equatorial coordinates compatible with low air masses for most of the night at the observing site;

• low or no presence of already known variables, where this element has been determined by analyzing the distribution of the variables in the International Variable Star Index operated by the AAVSO (VSX; Watson *et al.* 2014).

The observations were performed with the CCD at a temperature of –10° C (in summer and fall) and –20° C (in winter) in binning 1 × 1. The exposure time was 120 seconds with a delay of 1 second between the images and an average download time of 11 seconds per frame. Once the images were obtained, a full calibration procedure was performed with dark and flat frames. All images were then aligned and an astrometric reduction made to implement the astrometrical WCS coordinate system in the FITS header. All these operations were conducted entirely through the use of software MAXIMDL V5.23 (Diffraction Limited 2012). The complete observing log is reported in Table 1.

## 4. Methods and procedures

As a continuation of the wider survey activity conducted from the Keyhole Observatory MPC K48 for new variable star discovery, the methods and procedures used will be presented in a very schematic way; in previously published works (Furgoni 2013a, 2013b) more specific details can be found.

Table 1. Observing log.

*Vulpecula Field*
R.A. 19$^h$ 29$^m$ 00$^s$, Dec. +25° 50' 00"
(stars described in sections 5.1 through 5.7)

| UT starting time (yyyy/mm/dd hh.mm.ss) | UT ending time (yyyy/mm/dd hh.mm.ss) |
|---|---|
| 2014/07/03 20:30:45 | 2014/07/04 00:05:40 |
| 2014/07/15 20:24:53 | 2014/07/16 01:29:08 |
| 2014/07/16 20:19:39 | 2014/07/17 00:02:07 |
| 2014/07/17 20:17:06 | 2014/07/17 20:58:23 |
| 2014/08/07 19:49:18 | 2014/08/07 22:28:48 |
| 2014/08/18 19:27:24 | 2014/08/18 21:24:43 |
| 2014/08/23 21:11:15 | 2014/08/23 23:34:04 |
| 2014/08/28 19:10:18 | 2014/08/28 23:13:25 |
| 2014/09/14 18:32:17 | 2014/09/14 23:01:44 |
| 2014/09/16 18:46:49 | 2014/09/16 19:58:31 |
| 2014/09/22 18:37:41 | 2014/09/22 20:58:48 |
| 2014/09/25 18:13:26 | 2014/09/25 21:09:26 |
| 2014/09/27 18:38:34 | 2014/09/27 22:37:33 |
| 2014/10/24 17:21:46 | 2014/10/24 21:00:56 |
| 2014/10/26 17:20:31 | 2014/10/26 20:57:32 |

*Perseus 1 Field*
R.A. 03$^h$ 19$^m$ 00$^s$, Dec. +41° 29' 00"
(stars described in sections 5.8 through 5.12)

| UT starting time (yyyy/mm/dd hh.mm.ss) | UT ending time (yyyy/mm/dd hh.mm.ss) |
|---|---|
| 2015/01/11 17:28:35 | 2015/01/11 22:23:41 |
| 2015/01/12 17:14:17 | 2015/01/12 21:23:08 |
| 2015/01/26 17:23:42 | 2015/01/26 19:42:27 |
| 2015/01/28 17:23:44 | 2015/01/28 20:58:01 |
| 2015/02/02 17:50:58 | 2015/02/02 21:54:59 |
| 2015/02/09 17:45:05 | 2015/02/09 21:26:17 |

*Perseus 2 Field*
03$^h$ 19$^m$ 00$^s$, Dec. +42° 46' 00"
(stars described in sections 5.13 through 5.17)

| UT starting time (yyyy/mm/dd hh.mm.ss) | UT ending time (yyyy/mm/dd hh.mm.ss) |
|---|---|
| 2015/02/10 17:49:10 | 2012/02/10 22:23:19 |
| 2015/02/11 18:17:49 | 2015/02/11 22:03:50 |
| 2015/02/17 17:51:46 | 2015/02/17 21:28:15 |
| 2015/02/18 17:52:45 | 2015/02/18 21:34:48 |
| 2015/03/07 18:16:16 | 2015/03/07 21:47:15 |
| 2015/03/08 18:13:56 | 2015/03/08 21:37:38 |
| 2015/03/12 18:21:13 | 2015/03/12 21:36:25 |

The technique used for the determination of the magnitude was differential photometry, ensemble-type in almost all cases. It is necessary to have comparison stars with magnitudes accurately determined in order to obtain reliable measurements and as close as possible to the standard system. Since the entire observational campaign was conducted with the use of the Johnson V filter the comparison star magnitudes were obtained from the APASS (Henden *et al.* 2013) V magnitude data provided by the AAVSO.

The search for new variable stars was performed using the same technique described in Furgoni (2013a, 2013b): the method involves the distinction between discovery nights and follow-up nights. The first is needed to determine if the chosen field contains potential candidate variables through the creation of a magnitude-RMS diagram and the accurate visual inspection of all light curves down to the fourteenth magnitude. If in the discovery night the chosen field does not appear interesting it is discarded without further analysis. The follow-up nights, instead, are necessary for the collection of data concerning variable star candidates. In any case, any follow-up night is checked again by a magnitude-RMS scatter diagram. In this case, however, the light curves of the stars in the field are not inspected visually due to the long time needed for the operation. The combination of these two ways of working ensures a good compromise between the need to maximize the chance of discovery and to obtain the maximum from the data gradually collected.

Before proceeding further in the analysis, the time of the light curves obtained was heliocentrically corrected (HJD) in order to ensure a perfect compatibility of the data with observations carried out even at a considerable distance in time. When necessary, the determination of the period was calculated using the software PERIOD04 (Lenz and Breger 2005), using a Discrete Fourier Transform (DFT). The average zero-point (average magnitude of the object) was subtracted from the dataset to prevent the appearance of artifacts centered at frequency 0.0 of the periodogram. The calculation of the uncertainties was carried out with PERIOD04 using the method described in Breger *et al.* (1999).

To improve the period determination, SWASP, ASAS-3, and NSVS photometric data were used when available. However, due to the high scattering which in some cases affects both, the data with high uncertainties were eliminated. The datasets were also corrected in their zero-point to make them compatible with my V band data. Having the same zero-point is indeed crucial for correct calculation of the Discrete Fourier Transform operated by PERIOD04.

**5. New variable stars discovered**

In this survey seventeen new variable stars were discovered belonging to a population composed as follows:

• Three eclipsing (two β Lyrae and one β Persei).

• One eruptive (one RS Canum Venaticorum with β Persei eclipses).

• Ten pulsating (six Delta Scuti, one double-mode High Amplitude Delta Scuti, one RR Lyrae Ab-type, one RR Lyrae C-type and one Semi-Regular late-type).

• Three rotating (three rotating ellipsoidal, one of which is suspected).

The coordinates of all new variable stars discovered in this survey are reported as they appear in the UCAC4 catalogue (Zacharias *et al.* 2012) and never differ from the detected positions for a value greater than 1".

5.1. 2MASS J19305329+2558520
Position (UCAC4): R.A. (J2000) = 19$^h$ 30$^m$ 53.29$^s$, Dec. (J2000) = +25° 58' 52.0"
Cross Identification ID: UCAC4 580-084318
Variability Type: RR Lyr Ab
Magnitude: Max. 14.15 V, Min. 14.90 V
Period: 0.47303(2) d
Epoch: 2456898.3480(14) HJD
Rise duration: 19%
Ensemble Comparison Stars: UCAC4 580-083946 (APASS 11.479 V); UCAC4 580-083785 (APASS 11.665 V)
Check Star: UCAC4 580-084083
Notes: Phase plot is shown in Figure 1.

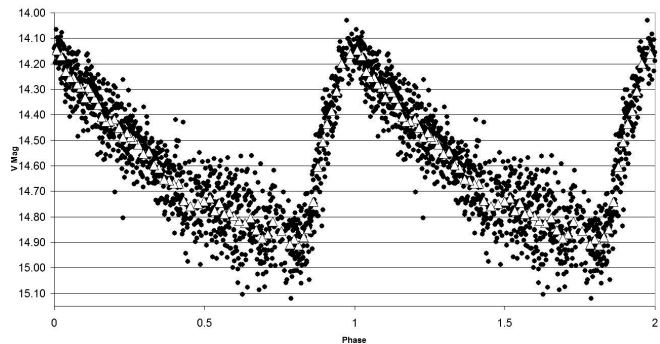

Figure 1. Phase Plot of 2MASS J19305329+2558520. Filled circles denote Furgoni data; open triangles denote Furgoni data (binning 20).

5.2. GSC 02129-00537
Position (UCAC4): R.A. (J2000) = 19$^h$ 31$^m$ 49.74$^s$, Dec, (J2000) = +25° 50' 49.8"
Cross Identification ID: 2MASS J19314974+2550498; UCAC4 580-084953
Variability Type: δ Sct
Magnitude: Max 13.42 V, Min 13.46 V
Period: 0.1039816(38) d
Epoch: 2456854.3844(11) HJD
Ensemble Comparison Stars: UCAC4 580-083946 (APASS 11.479 V); UCAC4 580-083785 (APASS 11.665 V)
Check Star: UCAC4 580-084083
Notes: Low galactic latitude star, probably reddened. Phase plot is shown in Figure 2.

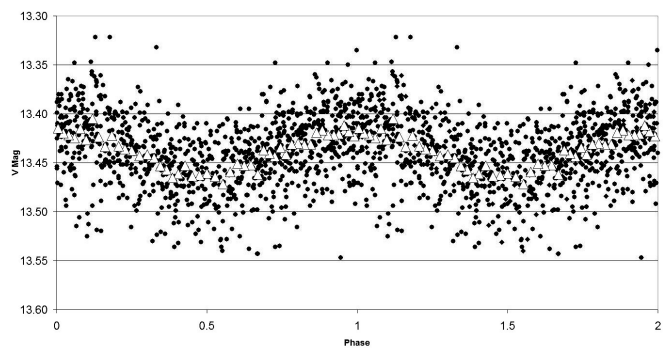

Figure 2. Phase Plot of GSC 02129-00537. Filled circles denote Furgoni data; open triangles denote Furgoni data (binning 20).

5.3. GSC 02142-01107
Position (UCAC4): R.A. (J2000) = 19$^h$ 32$^m$ 33.71$^s$, Dec. (J2000) = +25° 47' 42.7"
Cross Identification ID: 2MASS J19323371+2547427; ASAS J193233+2547.7; UCAC4 579-084405
Variability Type: Rotating ellipsoidal
Magnitude: Max 11.33 V, Min 11.42 V
Period: 1.243460(2) d
Epoch: 2456955.358(3) HJD
Ensemble Comparison Stars: UCAC4 580-083946 (APASS 11.479 V); UCAC4 580-083785 (APASS 11.665 V)
Check Star: UCAC4 580-084083
Notes: Secondary maximum = 11.37 V. Phase plot is shown in Figure 3.

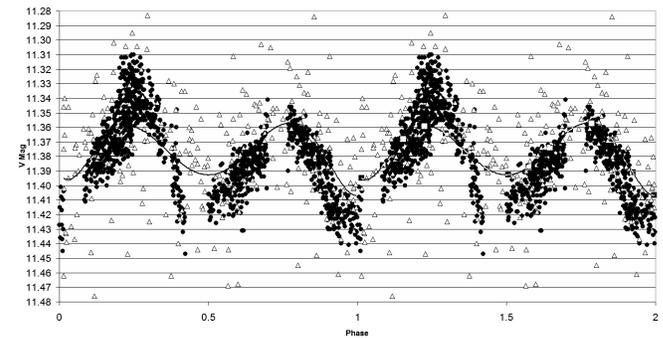

Figure 3. Phase Plot of GSC 02142-01107. Filled circles denote Furgoni data; open triangles denote ASAS3 data (–0.02 mag. offset); line denotes ASAS3 polynomial fitting.

5.4. GSC 02129-00759
Position (UCAC4): R.A. (J2000) = 19$^h$ 31$^m$ 13.21$^s$, Dec. (J2000) = +25° 43' 35.0"
Cross Identification ID: 2MASS J19311321+2543349; NSVS 8365171; UCAC4 579-083612
Variability Type: β Lyr
Magnitude: Max 13.55 V, Min 14.10 V
Period: 0.6647264(2) d
Epoch: 2456854.3911(25) HJD
Ensemble Comparison Stars: UCAC4 579-083181 (APASS 11.595 V); UCAC4 578-089176 (APASS 11.721 V)
Check Star: UCAC4 578-089646
Notes: Secondary minimum = 13.80 V with epoch 2456855.3882(25) HJD. Phase plot is shown in Figure 4.

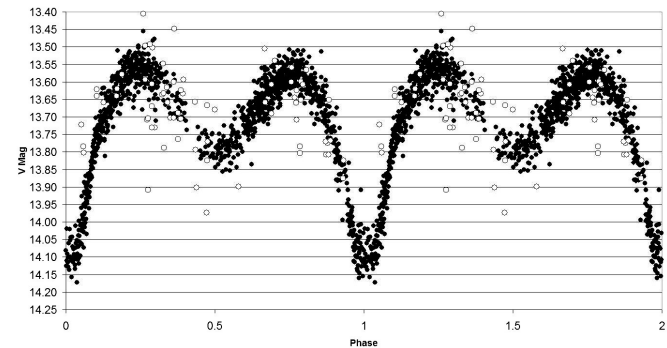

Figure 4. Phase Plot of GSC 02129-00759. Filled circles denote Furgoni data; open circles denote NSVS data (error < 0.3 mag.).

5.5. 2MASS J19323543+2524000
Position (UCAC4): R.A. (J2000) = $19^h 32^m 35.44^s$, Dec. (J2000) = +25° 24′ 00.1″
Cross Identification ID: UCAC4 578-090607
Variability Type: RR Lyr C
Magnitude: Max 13.67 V, Min 14.15 V
Period: 0.378192(6) d
Epoch: 2456915.330(1) HJD
Rise duration: 40%
Ensemble Comparison Stars: UCAC4 579-083181 (APASS 11.595 V); UCAC4 578-089176 (APASS 11.721 V)
Check Star: UCAC4 578-089646
Notes: Phase plot is shown in Figure 5.

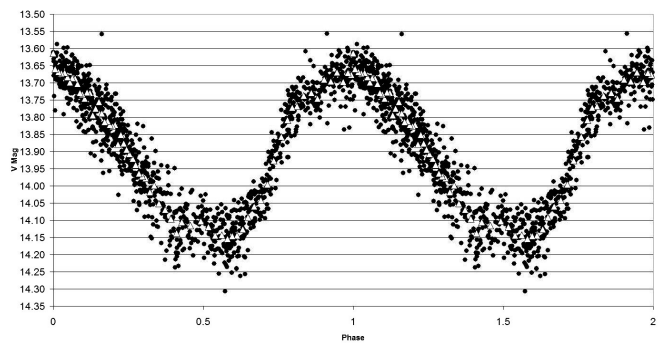

Figure 5. Phase Plot of 2MASS J19323543+2524000. Filled circles denote Furgoni data; open triangles denote Furgoni data (binning 20).

5.6. 2MASS J19263580+2616428
Position (UCAC4): R.A. (J2000) = $19^h 26^m 35.81^s$, Dec. (J2000) = +26° 16′ 42.8″
Cross Identification ID: UCAC4 582-080882; USNO-B1.0 1162-0380200
Variability Type: δ Sct
Magnitude: Max 13.76 V, Min 13.85 V
Period: 0.182787(7) d
Epoch: 2456854.4373(12) HJD
Ensemble Comparison Stars: UCAC4 580-081384 (APASS 11.856 V); UCAC4 580-082328 (APASS 12.230 V)
Check Star: UCAC4 580-082053
Notes: Phase plot is shown in Figure 6.

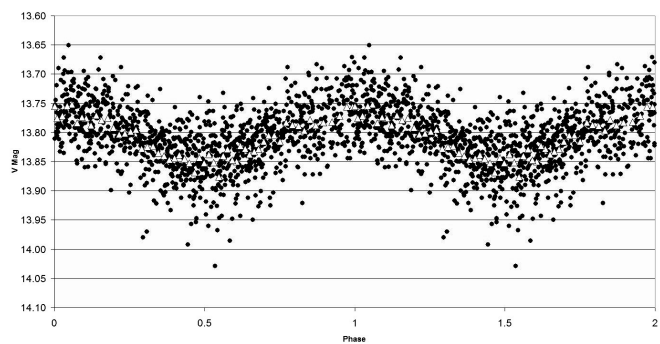

Figure 6. Phase Plot of 2MASS J19263580+2616428. Filled circles denote Furgoni data; open triangles denote Furgoni data (binning 20).

5.7. GSC 02129-00947
Position (UCAC4): R.A. (J2000) = $19^h 26^m 53.10^s$, Dec. (J2000) = +25° 52′ 00.1″
Cross Identification ID: 2MASS J19265309+2552000; NSVS 8359362; UCAC4 580-081821
Variability Type: β Per
Magnitude: Max 12.90 V, Min 13.30 V
Period: 0.6559882(2) d
Epoch: 2456898.342(2) HJD
Ensemble Comparison Stars: UCAC4 580-081384 (APASS 11.856 V); UCAC4 580-082328 (APASS 12.230 V)
Check Star: UCAC4 580-082053
Notes: Secondary minimum = 13.00 V with epoch 2456855.375(2) HJD. Phase plot is shown in Figure 7.

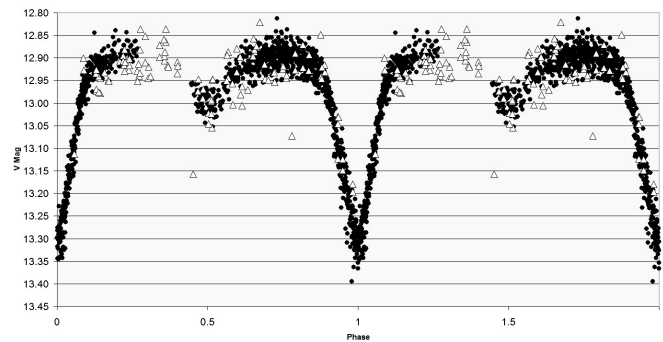

Figure 7. Phase Plot of GSC 02129-00947. Filled circles denote Furgoni data; open triangles denote NSVS data (error < 0.1 mag.; +0.18 mag. offset).

5.8. GSC 02856-02521
Position (UCAC4): R.A. (J2000) = $03^h 19^m 34.83^s$, Dec. (J2000) = +41° 56′ 04.8″
Cross Identification ID: 1RXH J031934.6+415606; 1SWASP J031934.83+415604.9; 2MASS J03193483+4156048; 3XMM J031934.8+41560; UCAC4 660-015379
Variability Type: RS CVn with β Per eclipses
Magnitude: Max 12.80 V, Min 13.09 V
Period: 3.820158(8) d
Epoch: 2457034.400(2) HJD
Ensemble Comparison Stars: UCAC4 660-015482 (APASS 11.188 V); UCAC4 659-015252 (APASS 11.243 V)
Check Star: UCAC4 660-015415
Notes: Eclipse duration 3%. Min II = 13.03 V. Slightly eccentric system with secondary minimum at phase 0.495 with epoch 2454056.568(2) HJD calculated from SWASP dataset. The system is an X-ray source identified both by the XMM-Newton Serendipitous Source Catalogue 3XMM-DR4 (XMM-Newton Survey Science Centre 2013) and the ROSAT Source Catalog of Pointed Observations with the High Resolution Imager (Rosat Scientific Team 2000) in the 0.2–2 KeV spectrum range. Phase plots are shown in Figures 8a–c; Figures 9a–c show XMM-Newton Source detection, time series, and X-ray spectrum.

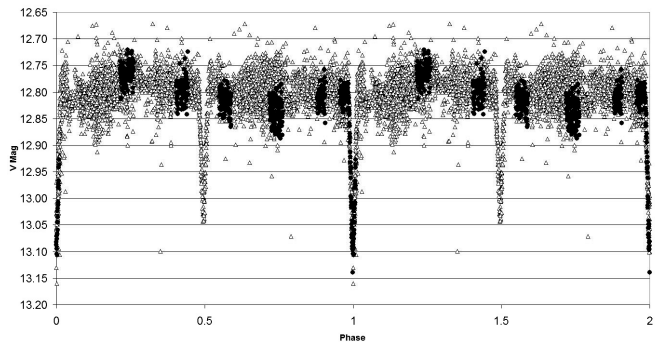

Figure 8a. Phase Plot of GSC 02856-02521. Filled circles denote Furgoni data; open triangles denote SWASP data (camera 142 and 141 with error < 0.03 mag.; +0.07 mag. offset).

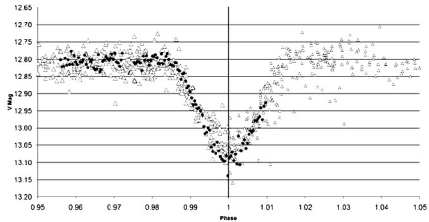

Figure 8b. Phase Plot of GSC 02856-02521. Zoom at phase 1.0.

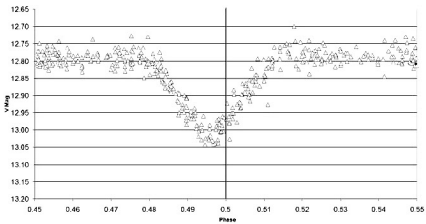

Figure 8c. Phase Plot of GSC 02856-02521. Zoom at phase 0.5.

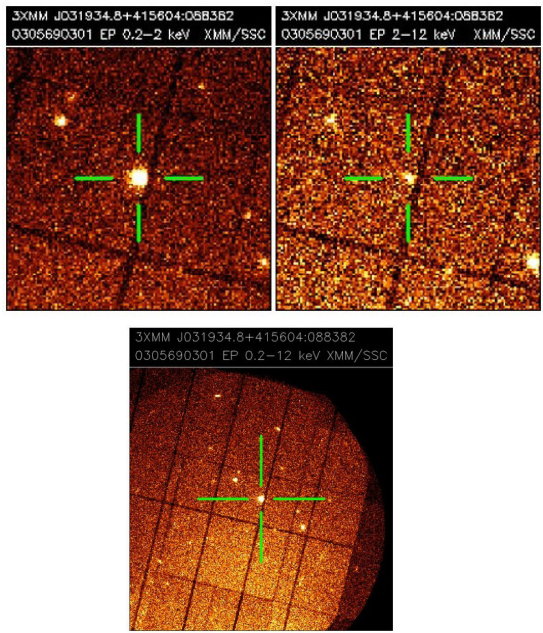

Figure 9a. XMM-Newton source detection of GSC 02856-02521.

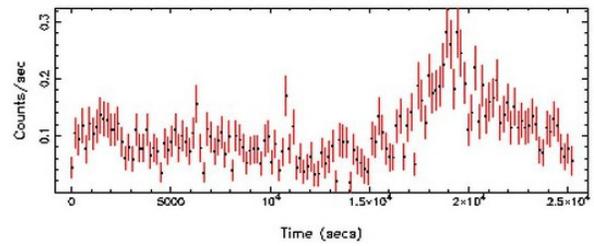

Figure 9b. Background subtracted time series of GSC 02856-02521; mean rate = 0.10366898.

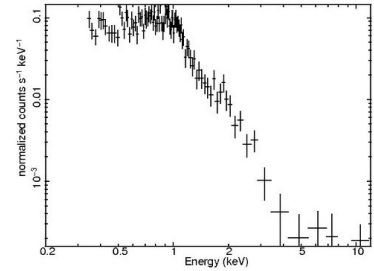

Figure 9c. X-ray spectrum of GSC 02856-02521. Srx 1 minus BGD (P0305690301 PNS003SRSPEC0001).

### 5.9. GSC 02865-01593

Position (UCAC4): R.A. (J2000) = $03^h 23^m 07.29^s$, Dec. (J2000) = +41° 14' 41.3"
Cross Identification ID: 2MASS J03230728+4114412; UCAC4 657-015233
Variability Type: Rotating ellipsoidal (suspected)
Magnitude: Max 11.78V, Min 11.80 V
Period: 0.4090484(18) d
Epoch: 2457034.4064(12) HJD
Ensemble Comparison Stars: UCAC4 656-015080 (APASS 11.141 V); UCAC4 656-015057 (APASS 11.740 V)
Check Star: UCAC4 657-015131
Notes: Phase plot is shown in Figure 10.

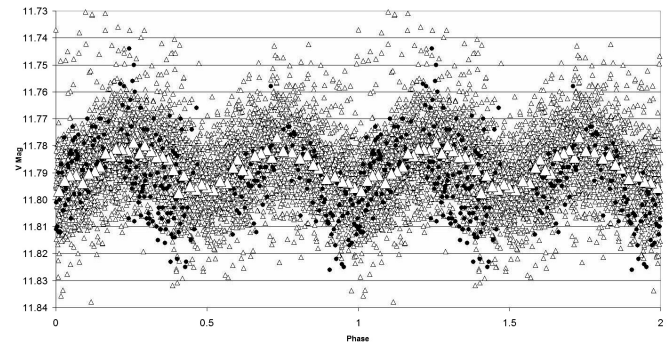

Figure 10. Phase Plot of GSC 02865-01593. Filled circles denote Furgoni data; small open triangles denote SWASP data (camera 148, 142, and 141 with error < 0.02 mag.; +0.26 mag. offset); large open triangles denote SWASP data (binning 40).

5.10. HD 275169

Position (UCAC4): R.A. (J2000) = 03$^h$ 21$^m$ 12.88$^s$, Dec. (J2000) = +40° 56' 45.6"

Cross Identification ID: 1SWASP J032112.86+405645.6; 2MASS J03211287+4056457; BD+40 718; GSC 02865-01872; TYC 2865-1872-1; UCAC4 655-014926

Variability Type: δ Sct

Magnitude: Max 10.61 V, Min 10.65V

Main Period: 0.09041008(6) d

Epoch of the Main Period: 2457034.3456(6) HJD

Ensemble Comparison Stars: UCAC4 656-015080 (APASS 11.141 V); UCAC4 656-015057 (APASS 11.740 V)

Check Star: UCAC4 657-015131

Notes: The star shows a clear variability of the light curve, where at least other two active frequencies were detected with the following elements: HJD 2457034.3442(7) + 0.10055449(8) E with SNR 4.1; HJD 2457034.330(2) + 0.1043577(2)E with SNR 3.0. Spectral type F0 from Nesterov *et al.* (1995). Phase plots and Fourier spectrum plots are shown in Figures 11a–f.

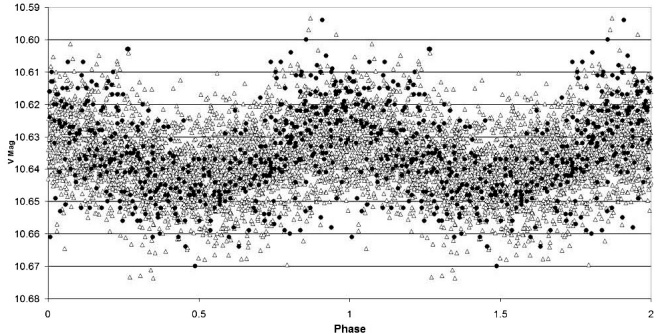

Figure 11a. Main period phase plot of HD 275169; period = 0.09041008(6) d; HDJ$_{max}$ = 2457034.3456(6). Filled circles denote Furgoni data; open triangles denote SWASP data (camera 148, 142, and 141 with error < 0.01 mag.; +0.11 mag. offset).

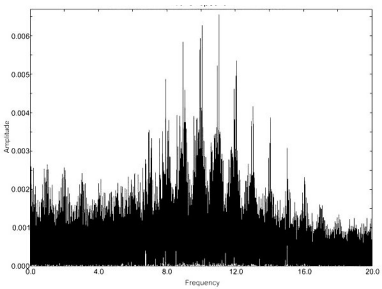

Figure 11b. Fourier spectrum of HD 275169; power spectrum.

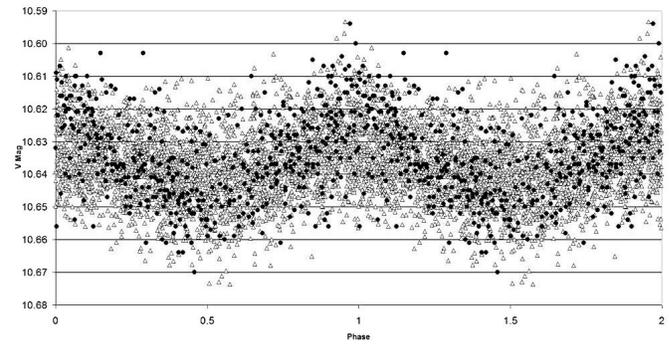

Figure 11c. Second period phase plot of HD 275169; period = 0.10055449(8) d; HDJ$_{max}$ = 2457034.3442(7). Filled circles denote Furgoni data; open triangles denote SWASP data (camera 148, 142, and 141 with error < 0.01 mag.; +0.11 mag. offset).

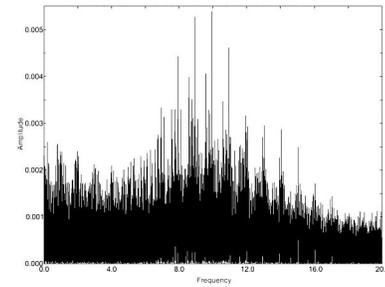

Figure 11d. Fourier spectrum of HD 275169 (after pre-whitening for the 11.060714 c/day frequency).

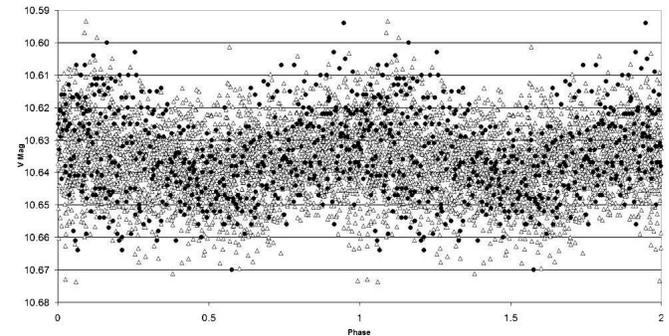

Figure 11e. Third period phase plot of HD 275169; period = 0.1042577(2) d; HDJ$_{max}$ = 2457034.330(2). Filled circles denote Furgoni data; open triangles denote SWASP data (camera 148, 142, and 141 with error < 0.01 mag.; +0.11 mag. offset).

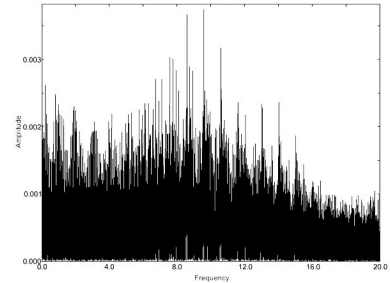

Figure 11f. Fourier spectrum of HD 275169 (after pre-whitening for the 11.060714 c/day 9.944856 c/day frequencies).

Furgoni,   *JAAVSO Volume 43, 2015*                                                                                                                        7

### 5.11. GSC 02869-00313

Position (UCAC4): R.A. (J2000) = 03$^h$ 22$^m$ 59.17$^s$, Dec. (J2000) = +41° 43' 40.5"
Cross Identification ID: 1SWASP J032259.30+414340.9; 2MASS J03225916+4143404; UCAC4 659-015509
Variability Type: δ Sct
Magnitude: Max 11.835 V, Min 11.845V
Period: 0.06505865(6) d
Epoch: 2457034.2432(7) HJD
Ensemble Comparison Stars: UCAC4 659-015465 (APASS 10.949 V); UCAC4 659-015381 (APASS 10.540 V)
Check Star: UCAC4 659-015447
Notes: Phase plot is shown in Figure 12.

### 5.12. GSC 02869-01981

Position (UCAC4): R.A. (J2000) = 03$^h$ 22$^m$ 38.08$^s$, Dec. (J2000) = +41° 49' 17.4"
Cross Identification ID: 1SWASP J032238.07+414917.5; 2MASS J03223807+4149173; AKARI-IRC-V1 J0322380+414917; TYC 2869-1981-1; UCAC4 660-015709
Variability Type: Semiregular late-type
Magnitude: Max 11.50 V, Min 11.65V
Period: 35.6(1) d
Epoch: 2457040.17(10) HJD
Ensemble Comparison Stars: UCAC4 659-015465 (APASS 10.949 V); UCAC4 659-015381 (APASS 10.540 V)
Check Star: UCAC4 659-015447
Notes: J–K = 1.223. The star is a MID-IR source (CDS catalog II/297/IRC) and probably also identified with the FAR-IR point source PSC 03192-4138 listed in Wise *et al.* (1993). Possibile presence of a secondary period. Phase plot is shown in Figure 13; light curve is shown in Figure 14.

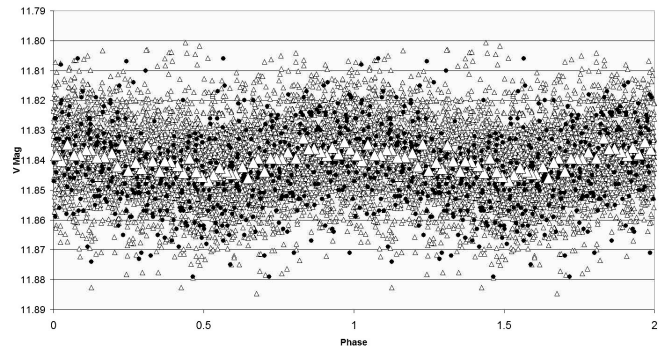

Figure 12. Phase Plot of GSC 02869-00313. Filled circles denote Furgoni data; small open triangles denote SWASP data (camera 141, 142, 144, and 148 with error < 0.02 mag.; –0.25 mag. offset); large open triangles denote SWASP data (binning 40).

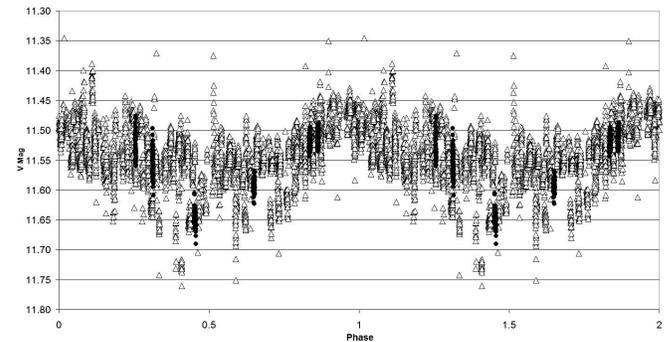

Figure 13. Phase Plot of GSC 02869-01981. Filled circles denote Furgoni data; open triangles denote SWASP data (camera 105, 141, 144, and 148 with error < 0.05 mag.; +0.45 mag. offset).

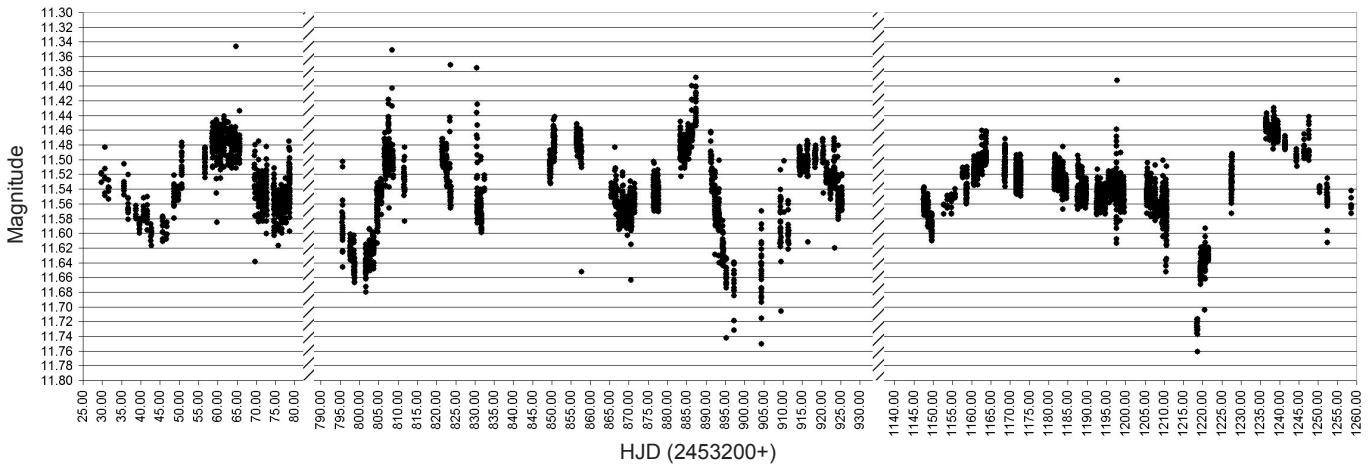

Figure 14. Light Curve of GSC 02869-01981. Y-axis = SWASP magnitude (only datapoint with < 0.05 mag. error and outliers rejected; +0.45 mag. offset); x-axis = HJD from 2453200.0).

5.13. GSC 02856-01391
Position (UCAC4): R.A. (J2000) = $03^h 19^m 35.51^s$, Dec. (J2000) = +43° 02' 41.3"
Cross Identification ID: 1SWASP J031935.48+430241.4; 2MASS J03193551+4302414; UCAC4 666-016419
Variability Type: δ Sct
Magnitude: Max 13.40 V, Min 13.47 V
Period: 0.1086275(7)
Epoch: 2457064.3295(9) HJD
Ensemble Comparison Stars: UCAC4 666-016582 (APASS 11.679 V); UCAC4 666-016538 (APASS 11.519 V)
Check Star: UCAC4 666-016446
Notes: J–K = 0.23; SWASP magnitudes contaminated by UCAC4 666-016412 (V = 11.500; sep. 24"). SWASP range has been corrected. Phase plot is shown in Figure 15.

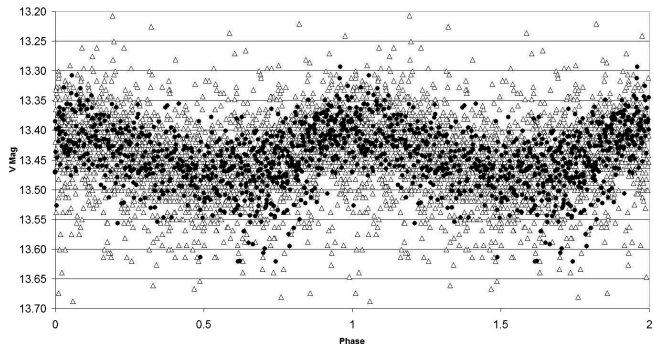

Figure 15. Phase Plot of GSC 02856-01391. Filled circles denote Furgoni data; open triangles denote SWASP data (camera 148, 142, and 141; 0.03 mag. offset and de-blending of UCAC4 666-016412).

5.14. GSC 02869-02559
Position (UCAC4): R.A. (J2000) = $03^h 22^m 22.08^s$, Dec. (J2000) = +42° 38' 09.7"
Cross Identification ID: 1SWASP J032222.06+423809.7; 2MASS J03222207+4238097; UCAC4 664-016648
Variability Type: β Lyr
Magnitude: Max 12.24 V, Min 12.42 V
Period: 0.8109062(3)
Epoch: 2457072.2259(12) HJD
Ensemble Comparison Stars: UCAC4 664-016588 (APASS 10.948 V); UCAC4 664-016596 (APASS 10.987 V)
Check Star: UCAC4 664-016695
Notes: J–K = 0.25; secondary maximum = 12.27 V; Min II = 12.36 V with epoch 2457065.333(2) HJD from the Furgoni dataset and 2453997.675(2) from the SWASP dataset. Phase plot is shown in Figure 16.

5.15. GSC 02860-01552
Position (UCAC4): R.A. (J2000) = $03^h 16^m 02.70^s$, Dec. (J2000) = +43° 20' 34.3"
Cross Identification ID: 1SWASP J031602.71+432034.4; 2MASS J03160269+4320342; UCAC4 667-016141
Variability Type: HADS Double Mode
Magnitude: Max 12.52 V, Min 13.02 V
Fundamental Period: 0.13831414(4)
Epoch Fundamental Period: 2457064.407(1) HJD
First Overtone Period: 0.10675322(2)
Epoch First Overtone: 2457064.317(1) HJD
Ensemble Comparison Stars: UCAC4 667-016142 (APASS 11.258 V); UCAC4 667-016207 (APASS 11.740 V)
Check Star: UCAC4 666-016112
Notes: The ratio between the fundamental and first overtone modes is 0.7718 and consistent with the Petersen diagram for double mode HADS with P0 close to 0.10 d. The amplitude of the first overtone is higher than that of the fundamental mode. The Fourier spectrum after prewhitening for F0 and F1 shows the existence of the coupling term (F2 = F1 + F0; F3 = F1 – F0; F4 = F2 + F3) (Table 2). Phase plot and Fourier spectrum are shown in Figures 17a–b and Figures 18a–b.

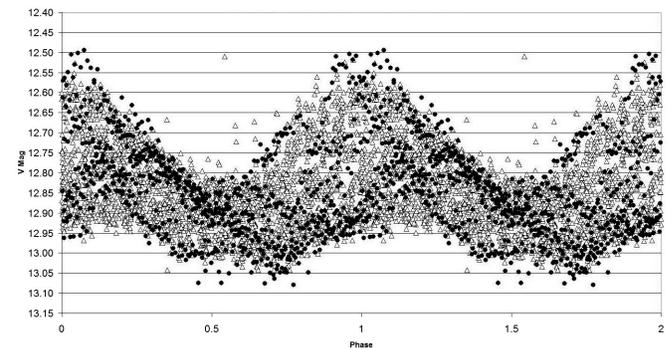

Figure 17a. Fundamental Mode Phase Plot of GSC 02860-01552. Filled circles denote Furgoni data; open triangles denote SWASP data (only camera 142 and 144 with error < 0.02 mag.; flux decontaminated from nearby stars).

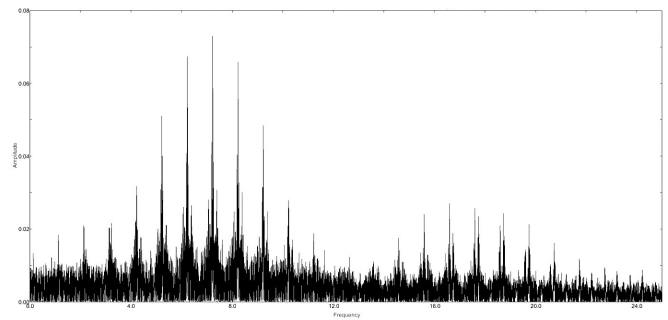

Figure 17b. Fourier Spectrum (after prewhitening for the first overtone frequency) of GSC 02860-01552.

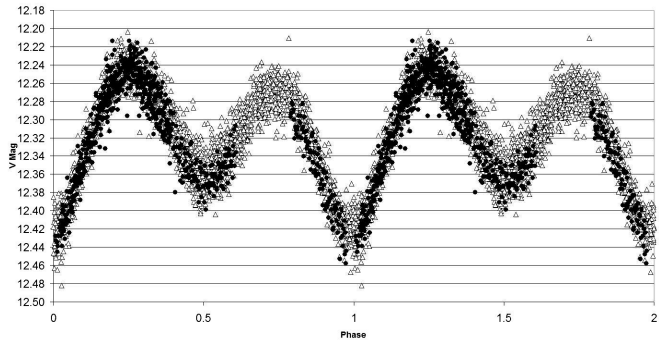

Figure 16. Phase Plot of GSC 02869-02559. Filled circles denote Furgoni data; open triangles denote SWASP data (only camera 142 and 148 with error < 0.02 mag.; –0.23 mag. offset).

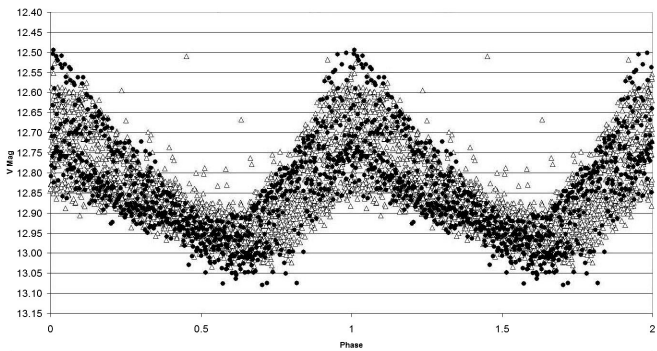

Figure 18a. First Overtone Phase Plot of GSC 02860-01552. Filled circles denote Furgoni data; open triangles denote SWASP data (only camera 142 and 144 with error < 0.02 mag.; lux decontaminated from nearby stars).

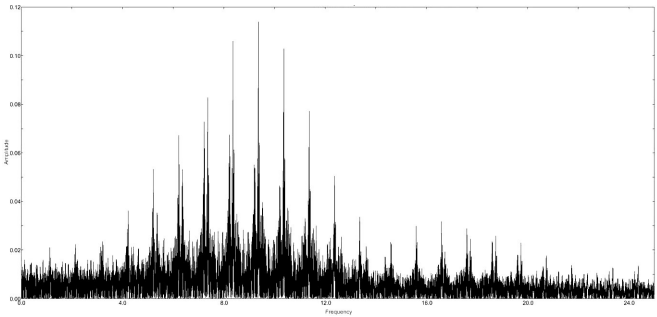

Figure 18b. Fourier Spectrum of GSC 02860-01552.

### 5.16. GSC 02856-01465

Position (UCAC4): R.A. (J2000) = $03^h 16^m 37.07^s$, Dec. (J2000) = +42° 40' 19.7"
Cross Identification ID: 1SWASP J031637.07+424019.8; 2MASS J03163707+4240196; UCAC4 664-015911
Variability Type: δ Sct
Magnitude: Max 12.70 V, Min 12.76V
Period: 0.12276888(5)
Epoch: 2457064.3240(3) HJD
Ensemble Comparison Stars: UCAC4 665-016194 (APASS 10.802 V); UCAC4 665-016217 (APASS 11.838 V)
Check Star: UCAC4 664-015818
Notes: The star is a reddened δ Sct: J–K = 0.182 and B–V = 0.269 after extinction correction based on Schlafly and Finkbeiner (2011). Phase plot is shown in Figure 19.

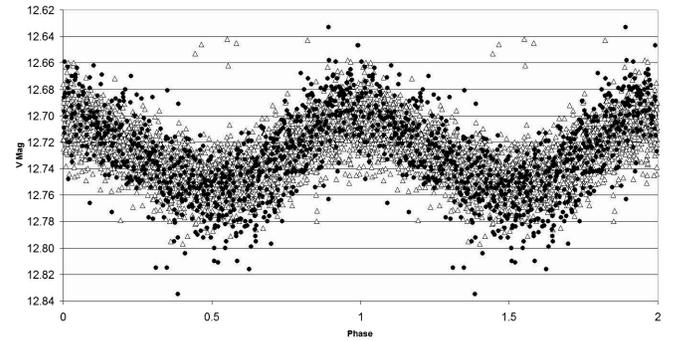

Figure 19. Phase Plot of GSC 02856-01465. Filled circles denote Furgoni data; open triangles denote SWASP data (only camera 142 and 144 with error < 0.02 mag.; +0.14 mag. offset).

Table 2. Coupling term elements for GSC 02860-01552.

| Frequency name | Frequency (c/d) | Amplitude | SNR | Period (d) | Err period (d) | $HJD_{max}$ | $HJD_{max}$ err |
|---|---|---|---|---|---|---|---|
| F0 | 7.22991898 | 0.074307369 | 6.39 | 0.13831414 | 3.95E-08 | 2457064.407 | 1.0E-03 |
| F1 | 9.36739897 | 0.115471575 | 9.86 | 0.10675322 | 1.51E-08 | 2457064.317 | 1.0E-03 |
| F2 (F1+F0) | 16.5973248 | 0.028539964 | 4.82 | 0.06025067 | 1.95E-08 | 2457064.416 | 2.0E-03 |
| F3 (F1-F0) | 2.13672248 | 0.023387893 | 3.10 | 0.46800650 | 1.44E-06 | 2457064.039 | 2.0E-03 |
| F4 (F2+F3) | 18.7355415 | 0.022394605 | 4.57 | 0.05337449 | 1.95E-08 | 2457064.423 | 3.0E-03 |

Table 3. Period calculations.

| Section (this paper) | Variable star | Epoch (HJD) | Period (d) |
|---|---|---|---|
| 5.3 | GSC 02142-01107 | 2456955.354(2) | 1.24345(4) |
| 5.4 | GSC 02129-00759 | 2456854.384(1) | 0.66471(4) |
| 5.7 | GSC 02129-00947 | 2456898.335(3) | 0.65600(3) |
| 5.9 | GSC 02865-01593 | 2457034.39(3) | 0.4087(5) |
| 5.10 | HD 275169 | 2457034.345(1) (Main Period) | 0.09040(2) (Main Period) |
| 5.11 | GSC 02869-00313 | 2457034.243(3) | 0.06509(9) |
| 5.13 | GSC 02856-01391 | 2457064.330(1) | 0.10862(1) |
| 5.14 | GSC 02869-02559 | 2457072.224(2) | 0.81091(2) |
| 5.15 | GSC 02860-01552 | 2457064.315(3) (First Overtone) | 0.106758(6) (First Overtone) |
|  |  | 2457064.406(2) (Fundamental) | 0.138301(9) (Fundamental) |
| 5.16 | GSC 02856-01465 | 2457064.324(1) | 0.122760(9) |
| 5.17 | GSC 02856-00169 | 2457072.327(1) | 0.28391(3) |

5.17. GSC 02856-00169

Position (UCAC4): R.A. (J2000) = 03$^h$ 16$^m$ 24.85$^s$, Dec. (J2000) = +42° 42' 55.4"
Cross Identification ID: 1SWASP J031624.90+424254.6; 2MASS J03162484+4242554; UCAC4 664-015890
Variability Type: Rotating Ellipsoidal
Magnitude: Max 13.80 V, Min 13.90 V
Period: 0.2838920(2)
Epoch: 2457072.3274(7) HJD
Ensemble Comparison Stars: UCAC4 665-016194 (APASS 10.802 V); UCAC4 665-016217 (APASS 11.838 V)
Check Star: UCAC4 664-015818
Notes: J–K = 0.553 and B–V = 0.805 after extinction correction based on Schlafly and Finkbeiner (2011). There is a 17.0 V mag. star 7" to the SSE. The magnitudes are for the blend of both stars. If the identification is correct and the variable is not the fainter star (unlikely), the deblended range would be V = 13.86–13.97.

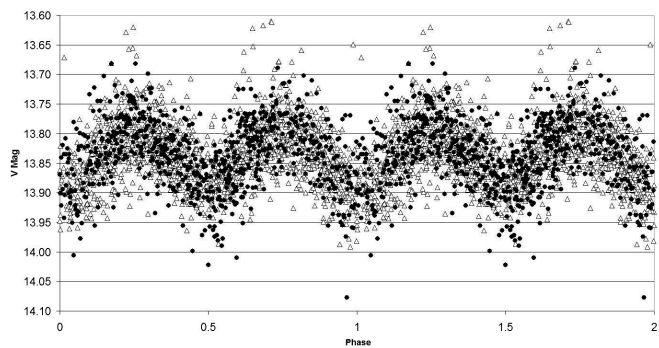

Figure 20. Phase Plot of GSC 02856-00169. Filled circles denote Furgoni data; open triangles denote SWASP data (only camera 142 and 144 with error < 0.04 mag.; –0.10 mag. offset).

**6. Ephemeris from Furgoni dataset**

Considering that the calculated epoch and period of almost all the variable stars discovered in this campaign are obtained from a combination of my observations and those from other surveys like ASAS, SWASP, and NSVS, it is necessary to point out that this way of work does not necessarily give precise period calculation. In effect almost all the periods are very short, and there are long lasting gaps between my observations and those in the other campaigns or databases. Therefore, in some cases there could be ambiguities in determining the accurate number of cycles that had passed during these gaps. Because of that I want to give for all the short period stars combined with other datasets an additional ephemeris calculated only with my dataset. However, for the stars listed in sections 5.8 and 5.12 I don't provide my own ephemeris because without the other survey datasets it is impossible to make a reasonable period calculation. In addition, for some stars where the cycle count problem is more relevant (sections 5.3, 5.4, 5.7, and 5.9) the period calculation was made weighting the data in relation to the uncertainty: with this approach I try to minimize the effect of most scattered data in the DFT calculation and provide a better period and error calculation. The results are presented in Table 3.

**7. Dedication and acknowledgements**

I want to dedicate this work to my loved cat Tillo, who passed away this summer, after many years as an irreplaceable companion under the starry sky.

This work has made use of the VizieR catalogue access tool, CDS, Strasbourg, France, and the International Variable Star Index (VSX) operated by the AAVSO, Cambridge, Massachusetts, USA. This work has made use of NASA's Astrophysics Data System and data products from the Two Micron All Sky Survey, which is a joint project of the University of Massachusetts and the Infrared Processing and Analysis Center/California Institute of Technology, funded by the National Aeronautics and Space Administration and the National Science Foundation.

This work has made use of the ASAS3 Public Catalog, NSVS data obtained from the Sky Database for Objects in Time-Domain operated by the Los Alamos National Laboratory, and data obtained from the SuperWASP Public Archive operated by the WASP consortium. This work has made use of observations obtained with XMM-Newton, an ESA science mission with instruments and contributions directly funded by ESA Member States and the USA (NASA).

This work has made use of The Fourth US Naval Observatory CCD Astrograph Catalog (UCAC4).